\documentclass[pra, aps, reprint]{revtex4-1}
\usepackage{bm}
\usepackage{graphicx}
\usepackage{siunitx}
\usepackage[final]{microtype}
\usepackage{amsmath, amsfonts, amssymb}
\usepackage{xcolor}
\usepackage{braket}
\newcommand{\ii}{\mathrm{i}}

\newcommand\diff{\mathop{}\!d}
\DeclareSIUnit\wn{\raiseto{-1}\cm}

\begin{document}

\title{Unified model of ultracold molecular collisions}

\author{James F. E. Croft}
\affiliation{The Dodd-Walls Centre for Photonic and Quantum Technologies, New Zealand}
\affiliation{Department of Physics, University of Otago, Dunedin, New Zealand}
\author{John L. Bohn}
\affiliation{JILA, NIST, and Department of Physics, University of Colorado, Boulder, Colorado 80309-0440, USA}
\author{Goulven Qu{\'e}m{\'e}ner}
\affiliation{Universit\'{e} Paris-Saclay, CNRS, Laboratoire Aim\'{e} Cotton, 91405, Orsay, France}

\begin{abstract}
A scattering  model is developed for ultracold molecular collisions, which allows inelastic processes,
chemical reactions, and complex formation to be treated in a unified way.
All these scattering processes and various combinations of them are possible
in ultracold molecular gases, and as such this model will allow the
rigorous parametrization of experimental results.
In addition we show how, once extracted, these parameters can be related to
the physical properties of the system, shedding light on fundamental aspects of
molecular collision dynamics.
\end{abstract}
\maketitle

\section{Introduction}\label{sec:intro}
Ultracold samples of molecules can be exquisitely controlled at the quantum
state level, allowing fundamental physical and chemical process to be studied
with unprecedented precision.
This control has been used
to study state-to-state chemistry with full quantum state resolution for all
reactants and products~\cite{wolf.dei.ea:state-to-state},
to probe the potential energy surface with exquisite
resolution~\cite{klein.shagam.ea:directly,yang.zhang.ea:observation},
and to study the role of nuclear spins in molecular
collisions~\cite{ospelkaus.ni.ea:quantum-state,kilaj.gao.ea:observation},
More recently an experiment has managed to probe the intermediate
complex of an ultracold reaction~\cite{hu.liu.ea:direct} as such it
is now possible to track the complete chemical process from reactants,
through intermediates, to products.

Understanding the fundamental physical and chemical process of ultracold
molecular collisions is also important because ultracold gases are fragile
systems prone to collisional processes that can transfer their atomic or
molecular constituents into untrapped states or else release large amounts of
kinetic energy, leading to trap loss and heating.
A new mechanism for loss in an ultracold molecular gas was
proposed~\cite{mayle.ruzic.ea:statistical,mayle.quemener.ea:scattering},
namely a half-collision process in which the reactant molecules share energy
in rotational and vibrational degrees of freedom, spending a long time lost in
resonant states of a four-body collision complex rather than promptly completing
the collision process.
This idea of transient complex formation, colloquially dubbed ``sticking'',
takes on an added significance for ultracold molecular collisions where the
number of available exit channels can be very small compared to the number of
resonant states.

Initial experiments on non-reactive ultracold molecules such as
NaRb~\cite{ye.guo.ea:collisions,guo.ye.ea:dipolar} and
RbCs~\cite{gregory.frye.ea:sticky} observed two-body collisional losses, even
though these species are non-reactive and are in their quantum mechanical
ground state so have no available inelastic loss channels.
As these complexes were not directly observed it remained an open question
whether these experiments have produced long-lived collisional complexes and
what the loss mechanism was.
However, a subsequent experiment on RbCs~\cite{gregory.blackmore.ea:loss}
showed that turning on or off the trapping light that confines the molecules
can increase or decrease the losses of the molecules.
This confirmed the hypothesis of a theoretical
study~\cite{christianen.zwierlein.ea:photoinduced} that the non-reactive molecules
first form tetramer complexes, and then the complexes are lost due to light
scattering in the optical dipole trap.
In addition, an experiment on chemically reactive ultracold molecules such as
KRb succeeded in directly observing the corresponding ions of the intermediate
complex K$_2$Rb$_2$~\cite{hu.liu.ea:direct}, as well as of the products K$
_2$ and Rb$_2$ of the chemical reaction.
Just as for non-reactive molecules the trapping light has a strong effect on
the losses of reactive molecules as well as on the lifetime of the transient
complex~\cite{liu.hu.ea:photo-excitation}, leading to the same conclusion
as~\cite{gregory.blackmore.ea:loss,christianen.zwierlein.ea:photoinduced}.
It is therefore clear that any theoretical treatment of ultracold molecular
collisions must be flexible enough to account for the formation of the complexes.

These experiments can be described by a model that assumes an absorption
probability $p_\mathrm{abs}$ for any two molecules that get within a certain
radius~\cite{idziaszek.julienne:universal}, without ascribing any particular
mechanism to the absorption.
Energy and electric field dependence of two-body loss rates are well-fit by the
resulting formulas.
For example, the reactive molecules in the KRb experiment vanish with unit
probability $p_\mathrm{abs}=1$ with or without electric
field~\cite{ospelkaus.ni.ea:quantum-state,ni.ospelkaus.ea:dipolar,de-marco.valtolina.ea:degenerate}.
The non-reactive species NaRb and RbCs
vanish with probabilities \num{0.89}~\cite{bai.li.ea:model} and
\num{0.66}~\cite{gregory.frye.ea:sticky} respectively, in zero electric field.
Notably, when an electric field is applied to NaRb, its absorption probability
climbs to $p_{\rm abs}=1$~\cite{guo.ye.ea:dipolar}.
Assuming the origin of this loss is due to complex formation, the increased
loss with electric field may be attributed to the increased density of
accessible states of the complex and/or coupling of these states to the
continuum scattering channels.
It is therefore conceivable that complex formation may be a phenomenon that can be
turned on or off as desired.

While the influence of light scattering on molecular collisions is undeniable,
it should be possible for the molecules to be confined in ``box'' traps where
the molecules remain mostly in the dark encountering trapping light only
at the peripheries of the trap~\cite{gaunt.schmidutz.ea:bose-einstein}.
In this case, loss due to complex formation would allow a more direct probe of
the fundamental four-body physics of the collision.

In this paper we propose a phenomenological model of collisional losses, based
on the theory of average cross sections~\cite{feshbach.porter.ea:model},
that encompasses both direct collisional losses and loss due to complex formation.
As such this model serves not only to parametrize experimental measurements,
but also allows those parameters to be related to the physical properties of
the system, potentially shedding light on the dynamics of the molecular complex.

\section{Theory}\label{sec:theory}
The theory must be flexible enough to describe the various outcomes available
when two molecules collide:
elastic scattering of the reactants; inelastic scattering, where
the reactants emerge with the same chemical identity but in different internal
states; reactive scattering into various product states; and absorption into
the collision complex.
Moreover, depending on the experiment, the various outcomes of the collision
may or may not be observed.
Note that within this model formation of a collision complex will always be
regarded an outcome in and of itself, we do not consider where the complex
ultimately decays to.

\subsection{Molecular scattering, observed and unobserved processes}\label{sec:molscat}
\begin{figure}[tb]
\centering
\includegraphics[width=0.99\columnwidth]{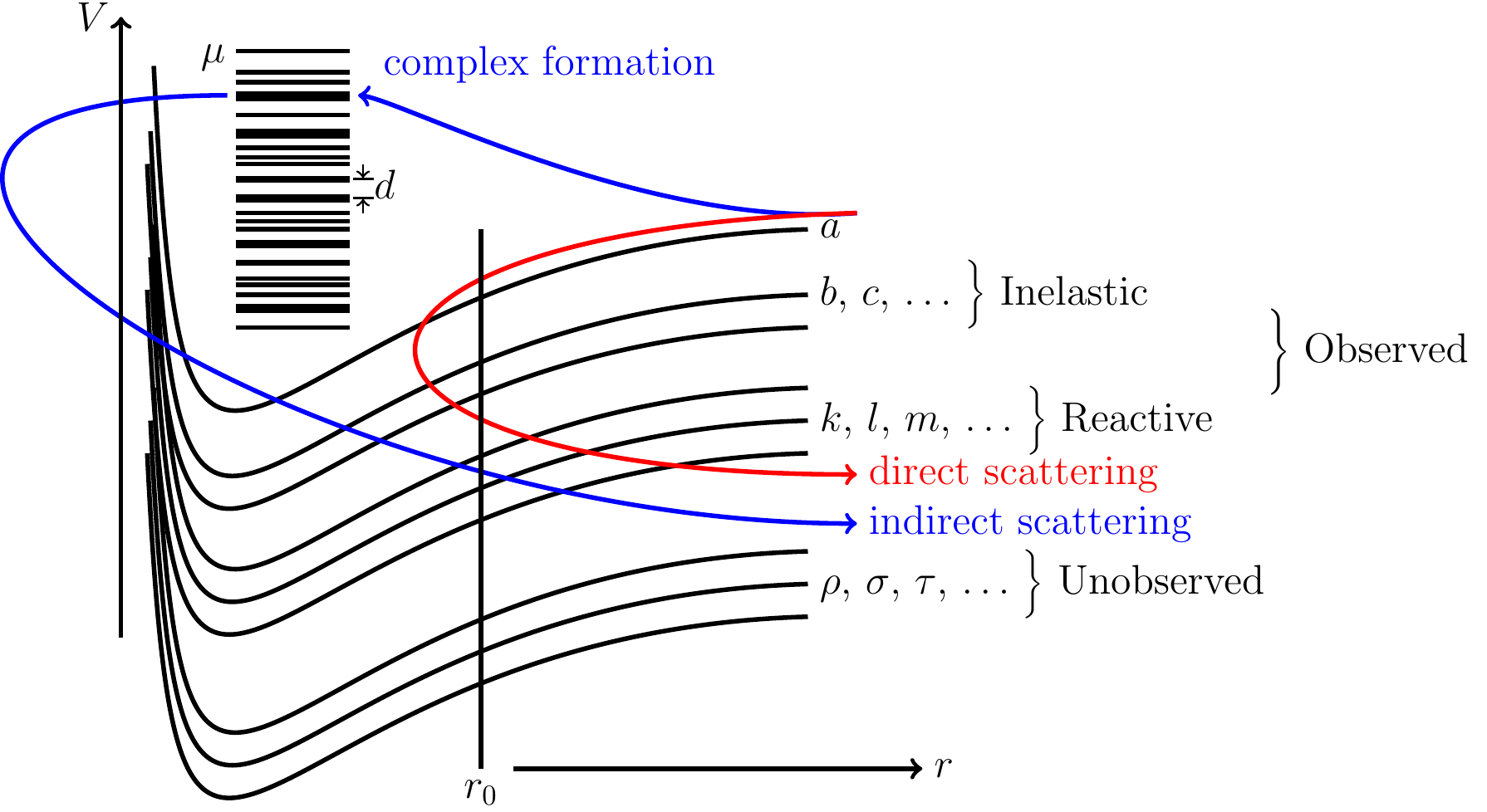}
\caption{Schematic figure showing potential energy curves $V$ versus
intermolecular distance $r$ (both in  arbitrary units), outlining the various
scattering processes and channel labels.
Direct and indirect scattering processes are shown by red and blue arrows
respectively.
The incoming channel is labelled $a$; inelastic scattering channels that can be
observed in an experiment are denoted by roman letters $b$, $c$, $\ldots$;
reactive scattering channels that can be observed in an experiment are denoted
by roman letters $k$, $l$, $m$, $\ldots$;
inelastic or reactive channels that are unobserved in a given experiment are
denoted by  letters $\rho$, $\sigma$, $\tau$, $\ldots$
Finally a dense forest of resonant states of the collision complex labelled by
$\mu$ with a mean level spacing $d$ (in arbitrary units), is shown in
the well of the potential.
}\label{fig:schematic}
\end{figure}

To this end, we define a flexible system of notation as illustrated in
Fig.~\ref{fig:schematic}.
This figure shows schematically the distance $r$ between two
collision partners (which may be reactants or products), and the various
possible outcome channels.
Channels whose outcome is observed by a particular experiment are labelled
by roman letters while channels whose outcome is unobserved are labelled by
greek letters.
The channels labelling observed processes are further differentiated as follows.
Channels labelled $a$, $b$, $c$, $\ldots$ correspond to the \emph{elastic} and
\emph{inelastic} channels of the reactants, while
channels labelled $k$, $l$, $m$, $\ldots$ correspond instead to channels of a
different molecular arrangement and correspond to the product channels of a
\emph{chemical reaction}.
Note that the unobserved processes may include both inelastic scattering as
well as chemical reactions: the criterion is simply that this outcome is not
observed.

By convention, we take $a$ to label the incident channel.
The list of channels $i$ will of course depend on the details of a particular
experiment.
In some experiments all the final states can be measured so that there are no
channels denoted by greek letters, while in others none of the final states can
be measured so that there are no channels denoted by roman letters
(except the incident channel $a$).
In some experiments inelastic channels are measured but reactive ones are not,
or vice versa.
Therefore one has to determine which processes are labelled as observed or
unobserved processes for a particular experiment of interest.
In Sec.~\ref{sec:avxs}, we will detail how these unobserved processes can be
gathered into an overall, absorption term.

The observed and unobserved processes are those that are expected to produce
inelastic scattering or chemical reactions immediately, that is, without forming
a collision complex, shown in Fig.~\ref{fig:schematic} by the red arrow
labelled \emph{direct scattering}.
Typically, the results of these processes release kinetic energy greater than
the depth of the trap holding the molecules, and hence lead to what we term
\emph{direct loss}.
By contrast, \emph{indirect scattering} processes which proceed via
\emph{complex formation}, shown in Fig.~\ref{fig:schematic} by the blue arrows,
will not immediately lead to trap loss.
Molecules lost into the collision complex require a second step to leave the
trap which consists of either absorbing a photon of trapping light,
colliding with another molecule, or decaying into an allowed channel of reactants
or products.
The present theory will not explicitly address this second step, focusing only
on the formation of the complex.

\subsection{Scattering cross sections for various processes}
A set of linearly independent scattering wavefunctions corresponding to these
various processes can be written in the channel state representation as
\begin{equation}
  \Psi_j = r^{-1}\sum_i \phi_i(\tau) F_{ij}(r),
\end{equation}
where $j$ runs up to the number of channels; the functions $\phi(\tau)$ form a
basis set for the motion in all coordinates, $\tau$, except the intermolecular
distance, $r$; and the elements $F_{ij}$ form a radial wave function matrix
$\bm F$~\cite{arthurs.dalgarno:theory,mies:scattering,quemener:ultracold}.
Applying the usual scattering boundary conditions at asymptotic separations
between the collision partners the radial wavefunction $\bm F$ defines, up to a
normalization, the scattering matrix $\bm S$
\begin{align}
\lim_{r \rightarrow \infty} \bm F \sim \psi^-(r) - \psi^+(r)\bm S,
\label{eq:S_matrix_def}
\end{align}
where $\psi^\pm$ are diagonal matrices with elements
\begin{align}
 \psi^{\pm}_i(r) = \frac{1}{\sqrt{k_i}} \exp(\pm\ii(k_i r - \pi l_i/2)),
\end{align}
and $k_i$ is the asymptotic wave vector and $l_i$ is the partial wave for
channel $i$.
Corresponding to each $S$-matrix element is the state-to-state probability of
a scattering process from channel $a$ to channel $i$,
with $i=a$, $b$, $c$, $\ldots$, $k$, $l$, $m$, $\ldots$, $\rho$, $\sigma$, $\tau$,
$\ldots$ given by
\begin{equation}
p_{a \to i}  =  \lvert S_{ia} \rvert^2.
\label{probasts}
\end{equation}
As the  $S$-matrix is unitary, we have for each incident channel $a$
\begin{align}
1&=\sum_{\substack{i=a, b, c, \ldots, \\ k, l, m, \ldots}} \!\!\! |S_{ia}|^2+ \! \sum_{i=\rho, \sigma, \tau, \ldots } \!\!\! |S_{ia}|^2 \nonumber \\
&\equiv p_\text{obs} + p_\text{unobs},
\label{Sunitarity}
\end{align}
where we have separated the scattering matrix into an observed block
($i$ running on Roman letters) and unobserved block
($i$ running on Greek letters).

The probability for observed processes, $p_\mathrm{obs}$, can be further
subdivided into elastic, inelastic and reactive parts
$p_\mathrm{obs} = p_\mathrm{el} + p_\mathrm{in} + p_\mathrm{re}$, with
\begin{align}
p_\mathrm{el} &=  p_{a \to a} =  \lvert S_{aa} \rvert^2 \nonumber \\
p_\mathrm{in} &= \!  \sum_{i=b, c, \ldots} \! p_{a \to i} = \sum_{i=b, c, \ldots} \! \lvert S_{ia}\rvert^2 \nonumber \\
p_\mathrm{re} &= \!\!\! \sum_{i=k, l, m, \ldots} \!\!\!\! p_{a \to i} = \!\!\! \sum_{i=k, l, m, \ldots} \!\!\! \lvert S_{ia} \rvert^2.
\label{probaelinrequ}
\end{align}
There is of course no reason to subdivide unobserved processes in this way.

In general, if some processes are unobserved,  then from Eq.~\eqref{Sunitarity},
the observed block of the scattering matrix will appear sub-unitary:
\begin{equation}
\sum_{i=a, b, c, \ldots, k, l, m, \ldots} \mkern-35mu |S_{ia}|^2 \leq 1.
\end{equation}
The amount by which this sum falls short of unity will be a measure of the
unobserved probability, which in general is recast into an overall absorption
probability $p_\mathrm{abs}$.
Therefore
\begin{equation}
p_\mathrm{abs} \equiv  p_\mathrm{unobs} = 1 - p_\mathrm{obs} =
1 - p_\mathrm{el} - p_\mathrm{in} - p_\mathrm{re}.
\label{probaabs}
\end{equation}
In the next subsection we will see that collisions resulting in complex
formation can also be properly included in the absorption probability.
It is also convenient to define a quenching probability which is the sum of
the inelastic, reactive, and absorption probabilities
\begin{equation}
p_\mathrm{qu} =  p_\mathrm{in} + p_\mathrm{re} + p_\mathrm{abs} = 1 - p_\mathrm{el} = 1 - \lvert S_{aa}\rvert^2.
\end{equation}
Finally, the corresponding state-to-state cross section is given by
\begin{equation}
\sigma_{a \to i} = g \frac{\pi}{k^2} \lvert\delta_{ai} - S_{ia} \rvert^2,
\end{equation}
where $g=2$ if the identical collision partners are initially in the same
quantum mechanical state and 1 otherwise.
The cross sections corresponding to these processes are given by
\begin{align}
  \sigma_\mathrm{el} &=  g \frac{\pi}{k^2} \lvert 1 - S_{aa}\rvert^2, \nonumber \\
  \sigma_\mathrm{in} &=  g \frac{\pi}{k^2} \ p_\mathrm{in}, \nonumber \\
  \sigma_\mathrm{re} &= g \frac{\pi}{k^2} \ p_\mathrm{re}, \nonumber \\
  \sigma_\mathrm{abs} &=  g \frac{\pi}{k^2} \ p_\mathrm{abs},  \nonumber \\
  \sigma_\mathrm{qu} &= \sigma_\mathrm{in} + \sigma_\mathrm{re} +  \sigma_\mathrm{abs}.
\label{XSusual}
\end{align}

\subsection{Complex formation and highly resonant collisions}\label{sec:highreso}
The treatment so far has ignored the possibility of complex formation.
The collision complex is comprised of a dense forest of resonant states,
depicted schematically in Fig.~\ref{fig:schematic} by
horizontal lines.
These states, denoted $\mu$, are potentially numerous and complicated and are
therefore best characterized by statistical quantities such as their mean
level spacing $d$ (equivalently mean density of states $\rho = 1/d$)
and their coupling matrix elements $W_{i \mu}$ to the open channels.
For the purposes of the theory it is assumed that the lifetime of the complex
is long compared to the mean collision time, so that complex formation and
decay can be considered distinct events.
This assumption appears to be validated by the observation of
K$_2$Rb$_2$ complexes in KRb + KRb ultracold reactions~\cite{hu.liu.ea:direct}
and by the measure of their lifetimes~\cite{liu.hu.ea:photo-excitation}.
In this case, losses due to complex formation can also occur.
Complex formation can therefore be considered as an unobserved process, as
defined in the previous section.
We will see that the theory of highly resonant collisions can recast this type
of loss into a phenomenological absorption term, included in the definition
(\ref{probaabs}) for the unobserved processes.

The key concept for understanding highly resonant collisions and the
corresponding complex formation can be found in the compound nucleus (CN) model
introduced by Bohr for understanding nuclear collisions~\cite{bohr:neutron}.
This theory postulates that a compound state involving all the nucleons forms
immediately when a particle (such as a neutron) encounters the nucleus.
These compound states have long lifetimes which leads to a dense set of narrow
resonances in the cross section as a function of energy.
Since Bohr's initial insight, the statistical theory of highly resonant
scattering has been developed considerably~\cite{feshbach.weisskopf:schematic,
weisskopf:compound,feshbach.porter.ea:model,friedman.weisskopf:compound,
feshbach:optical,feshbach:unified,feshbach:unified*1}.
We draw heavily on this literature in what follows.

The essential simplification of the statistical theory is the assumption that the
density of states is too great for any of the individual resonances to be
resolved, therefore scattering observables can  be replaced by suitable
averages~\cite{feshbach.porter.ea:model,friedman.weisskopf:compound}.
The virtue of this approach can be illustrated using an example with a
single channel, where the cross sections for elastic and absorption scattering
are given by
\begin{align}
 & \sigma_\mathrm{el} = g \frac{\pi}{k^2}\lvert1-S\rvert^2
  & \sigma_\mathrm{abs} =  g\frac{\pi}{k^2}{(1-\lvert S\rvert^2)},
 \label{eqn:xs_el_abs}
\end{align}
the overall absorption process accounting for all but the elastic process,
similar to what can be seen in Eq.~\eqref{XSusual} and Eq.~\eqref{probaabs}.
If all the incident flux were reflected then $|S|^2=1$ and there would be
no absorption.
If only a part of the flux were reflected $|S|^2$ would be less than unity,
due to absorption.
For indirect processes like complex formation there is no true absorption,
eventually the molecules re-emerge and complete a scattering event.
However, long-lived complexes can give the appearance of absorption
if the lifetime of the complexes is long enough, and moreover leads to true loss
if the complex is destroyed by a photon or by a collision with another molecule.
These effects are accounted for in the following.
The lifetime and subsequent decay of the complex is not treated within the
model we detail here.

In the statistical theory of scattering, the average of any energy-dependent
quantity $f(E)$ can be defined as
\begin{align}
\langle f(E) \rangle = \frac{1}{Z} \int \diff\epsilon \, f(\epsilon) \, D(E;\epsilon),
\end{align}
where $D(E;\epsilon)$ is the distribution, centered at $E$, that defines the average,
and $Z = \int \diff\epsilon \,  D(E;\epsilon)$.
$D$ is often taken to be either a Lorentzian function or else a finite step
function centered at $E$.
In any event, here $D$ is assumed to be broad enough to contain many resonances.
In these terms and factoring out the explicit momentum dependence,
an average cross section can be written~\cite{friedman.weisskopf:compound}
\begin{equation}
 \braket{\sigma} = g\frac{\pi }{k^2}\braket{k^2\sigma}.
\end{equation}
To calculate the average elastic cross section then requires taking the average
$\braket{|1 - S|^2}$.
Using the definition for the variance for a variable $X$
\begin{equation}
\Delta X \equiv \braket{\lvert X\rvert^2} - \lvert \braket X\rvert^2,
\label{eqn:var}
\end{equation}
we obtain
\begin{equation}
  \braket{|1-S|^2} = \lvert 1 - \braket{S} \rvert^2 + \Delta S.
\end{equation}
The average elastic cross section can therefore be written in the form
\begin{align}
\label{eqn:xs_avg_el}
\braket{\sigma_\mathrm{el}} &= g\frac{\pi}{k^2} \braket{|1 - S|^2}  \nonumber \\
  &=  g\frac{\pi}{k^2}{\lvert 1-\braket{S} \rvert^2} + g\frac{\pi}{k^2} \, \Delta S \nonumber \\
  &\equiv  \sigma_\mathrm{se} + \sigma_\mathrm{ce}.
\end{align}
These two contributions comprise a mean cross section, denoted the
``shape elastic'' cross section; and a contribution from the fluctuations,
denoted the ``compound elastic'' cross section~\cite{feshbach.porter.ea:model,
friedman.weisskopf:compound}.
Since the lifetime of a collisional process is proportional to the energy
derivative of the $S$-matrix~\cite{eisenbud:formal,wigner:lower,smith:lifetime,
frye.hutson:time},
writing the cross section in this way elegantly separates out the cross section
for direct scattering, the shape elastic part, from indirect scattering,
the compound elastic part.
Meanwhile, the average absorption cross section is
\begin{equation}
\label{eqn:xs_avg_abs}
 \braket{\sigma_\mathrm{abs}}=   g \frac{\pi}{k^2}(1-\braket{\lvert S\rvert^2}).
\end{equation}

The essence of the CN model is to associate elastic scattering with just the
shape elastic part of the elastic cross section and include the compound elastic
part in the absorption cross section.
This is achieved by simply making the replacement $S \to \braket{S}$ in
Eq.~\eqref{eqn:xs_el_abs}.
The elastic cross section in the CN model is therefore
\begin{align}
  \tilde\sigma_\mathrm{el} &= g\frac{\pi}{k^2}\lvert 1-\braket{S} \rvert^2 = \sigma_\mathrm{se}.
  \label{eq:CN_elastic}
\end{align}
while the absorption cross section is
\begin{align}
  \tilde\sigma_\mathrm{abs} &= g\frac{\pi}{k^2}{(1-\lvert\braket S\rvert^2)}, \nonumber \\
    &= g\frac{\pi}{k^2}(1-\braket{\lvert S\rvert^2})+g\frac{\pi}{k^2} \, \Delta S \nonumber \\
    &=  \braket{\sigma_\mathrm{abs}} + {\sigma}_\mathrm{ce}.
  \label{eq:CN_abs}
\end{align}
again using the definition of the variance.
As desired, simply by replacing $S$ in Eq.~\eqref{eqn:xs_el_abs} with
$\braket{S}$, the compound elastic part now appears in the absorption
cross section.

\subsection{Generalized theory of average cross sections}\label{sec:avxs}
Generalizing the averaging procedure above to the
multichannel case, one incorporates the effect of resonant complex formation by
energy averaging the appropriate cross sections over many resonances \cite{mitchell.richter.ea:random},
\begin{align}\label{probasts-dev}
 \braket{\sigma_{a \to i}}  &=  g\frac{\pi}{k^2}\braket{\lvert S_{ia} \rvert^2} \nonumber \\
  &= g\frac{\pi}{k^2}\lvert \braket{S_{ia}} \rvert^2  +  g\frac{\pi}{k^2}\Delta S_{ia} \nonumber \\
  &\equiv  \sigma_{a \to i}^\mathrm{dir} + \sigma_{a \to i}^\mathrm{ind}.
\end{align}
Doing so defines two components of the scattering.
The first component associated with an energy-smooth $S$-matrix $\braket{\bm S}$
is \emph{direct scattering}, which is scattering from channel $a$ to $i$ which proceeds without
forming a collision complex---this is the generalization of the
``shape elastic'' cross section in the single channel example above.
The second component associated with the energy fluctuations of the $S$-matrix $\Delta S$
is \emph{indirect scattering}, which is also scattering from channel $a$ to $i$ but
which proceeds via a collision complex---this is the generalization of the
``compound elastic'' cross section in the single channel example above.
As such this approach is completely general and can equally well
treat systems which proceed purely directly, purely indirectly, or any
combination thereof.
For ultracold collisions of alkali dimers, which we are primarily interested
in here, the potential energy surface is barrierless meaning in principle there
is no reason why direct reactions shouldn't
occur~\cite{ospelkaus.ni.ea:quantum-state,byrd.montgomery.ea:structure,
balakrishnan:perspective}.

More insight can be gained into the distinction between direct and
indirect processes by considering the time behaviour of an incident wave
packet~\cite{friedman.weisskopf:compound}, which follows directly from the
definition of the cross sections in Eq.~(\ref{probasts-dev}).
$\sigma^\mathrm{dir}$ is defined to vary slowly with energy, whereas
$\sigma^\mathrm{ind}$ contains all the rapid variation in the
cross section due to any resonances.
The component of the wave packet associated with direct scattering therefore
has a time behaviour similar to a scattering problem on the same potential in
the absence of any resonances.
On the other hand the component of the wave packet associated with indirect
scattering has a time behaviour similar to the long lifetime of a resonant
state.
As such the indirect component of the wave packet will come out delayed
compared to the direct component.

Following the prescription of the CN model, the cross sections for any process
whose outcome is observed are associated with the corresponding direct scattering
cross section.
The elastic, inelastic, and reactive cross sections are therefore given by
\begin{align}
\tilde\sigma_\mathrm{el} &\equiv \sigma_\mathrm{el}^\mathrm{dir} = g\frac{\pi}{k^2}{\lvert 1-\braket{S_{aa}} \rvert^2}   \nonumber \\
\tilde\sigma_\mathrm{in} &\equiv \sigma_\mathrm{in}^\mathrm{dir} = g\frac{\pi}{k^2} \! \sum_{i=b, c, \ldots} \! \lvert \braket{S_{ia}} \rvert^2    \nonumber \\
\tilde\sigma_\mathrm{re} &\equiv \sigma_\mathrm{re}^\mathrm{dir} = g\frac{\pi}{k^2} \!\!\!\! \sum_{i=k, l, m, \ldots} \!\!\!\! \lvert \braket{S_{ia}} \rvert^2.
\end{align}
These are presented as total cross sections,
for example, $\tilde\sigma_\mathrm{in}$ is the total inelastic cross section
and includes all inelastic scattering of molecules that is observed.
If individual inelastic channels are resolved experimentally, they correspond
to individual terms of this sum; and the same for reactive scattering.

Direct processes to channels that are not observed contribute to the absorption
cross section, $\tilde\sigma_\mathrm{abs}$, and their total contribution is
formally given by
\begin{equation}
 \tilde\sigma_\mathrm{abs}^\mathrm{dir} \equiv \sigma_\mathrm{abs}^\mathrm{dir} = g\frac{\pi}{k^2} \!\!\! \sum_{i=\rho, \sigma, \tau, \ldots} \!\!\! \lvert\braket{S_{ia}}\rvert^2.
\end{equation}
The cross section for complex formation also contributes to the absorption
cross section, and is simply the total cross section for all indirect
processes.
As such it is given by
\begin{align}
  \tilde\sigma_\mathrm{abs}^\mathrm{ind} & \equiv \sigma_\mathrm{el}^\mathrm{ind} + \sigma_\mathrm{in}^\mathrm{ind} + \sigma_\mathrm{re}^\mathrm{ind} + \sigma_\mathrm{abs}^\mathrm{ind} \nonumber \\
& =  g\frac{\pi}{k^2} \!\!\! \sum_{\substack{i=a, b, c, \ldots, \\  k, l, m, \ldots, \\ \rho, \sigma, \tau, \ldots}} \!\!\!\! \Delta S_{ia}.
\end{align}
The total absorption cross section is then the sum of the direct and the
indirect contributions
\begin{equation}
\tilde\sigma_\mathrm{abs} = \tilde\sigma_\mathrm{abs}^\mathrm{dir} + \tilde\sigma_\mathrm{abs}^\mathrm{ind}.
\label{XS-tilde}
\end{equation}
This is the generalization of Eq.~\eqref{eq:CN_abs}.
In this theory, the matrix elements of the highly-resonant $S$-matrix $S_{ia}$
(and therefore also $\Delta S_{ia}$) are presumed to be unknown.
Information about absorptive scattering will be inferred from the sub-unitarity
of the energy averaged $S$-matrix, $\braket{\bm{S}}$, in the observed channels.
Appropriate forms of these effective $S$-matrices will be derived in
Sec.~\ref{sec:absproc}.

It should be emphasized that by treating the complex as an absorption process
we are describing only the phenomenon of molecules combining to form a
collision complex.
Eventually such a complex would decay producing outcomes in any available
channel, but a full treatment of this process would require detailed
understanding of the decay mechanism of the complex, or equivalently the full
$S$-matrix on a fine enough energy grid that the appropriate averages given
in Eq.~(\eqref{probasts-dev}) could be meaningfully performed.

Finally, it is often useful to define a quenching cross section,
which describes scattering into any channel other than the incident channel,
regardless of what that channel is.
The quenching cross section is therefore given by
\begin{equation}
\tilde\sigma_\mathrm{qu} = \tilde\sigma_\mathrm{in} +\tilde\sigma_\mathrm{re}  +\tilde\sigma_\mathrm{abs} = g\frac{\pi}{k^2}(1- \lvert \braket{S_{aa}} \rvert^2).
\label{XS-tilde-qu}
\end{equation}

The corresponding rate coefficients $\tilde{\beta}$ to the cross sections given
above are obtained by replacing $1/k^2$ by $\hbar / \mu k$ where $\mu$ is the
reduced mass of the colliding molecules.

Hereafter, we adopt the perspective of CN theory: $S$-matrices will be
averaged over many resonances, and the resulting mean values will be used to
evaluate cross sections.

\subsection{Threshold Behavior}\label{sec:mqdt}
To facilitate applying the CN model to ultracold molecular scattering it is
useful to separate the effects of averaging and absorption from threshold
effects due to low collision energies.
The $S$-matrix defined in Eq.~\eqref{eq:S_matrix_def} is, in general, energy
dependent which leads to the usual Bethe-Wigner threshold laws for the
cross section~\cite{bethe:theory,wigner:on}.
This energy dependence is unrelated to the microscopic interactions between the
colliding molecules at small $r$ that dictate the molecular scattering processes.
As such these processes are better parametrized by energy-independent
short-range quantities.
To do so, we employ the ideas and methods of Multichannel Quantum Defect
Theory (MQDT)~\cite{seaton:quantum*1,fano:unified,greene.fano.ea:general,
greene.rau.ea:general,mies:multichannel,mies.julienne:multichannel,
burke.greene.ea:multichannel,gao:quantum-defect,mies.raoult:analysis,gao:general,
croft.wallis.ea:multichannel,ruzic.greene.ea:quantum,jachymski.krych.ea:quantum-defect},
which has been successfully applied in various ways and with various notations
to the problem of ultracold scattering.
At present, the version of this theory most commonly applied in ultracold
collisions is the Mies-Julienne version, whose notation  we follow
here~\cite{mies:multichannel,mies.julienne:multichannel}.

Scattering theory is usefully described in terms of real-valued, asymptotic
reference functions in each channel.
These functions are solutions to a single-channel Schr\"odinger equation, with
some predetermined potential.
They therefore do not represent free plane waves, but possesses a phase shift
$\xi_i$, their asymptotic form is
\begin{align}
f_i(r) &= \frac{1}{\sqrt{k_i}}\sin(k_i r - \pi l_i/2 + \xi_i) \nonumber \\
g_i(r) &= \frac{1}{\sqrt{k_i}}\cos(k_i r - \pi l_i/2 + \xi_i).
\end{align}
In terms of these functions, the asymptotic wave function can be written
\begin{equation}\label{eqn:react}
\lim_{r \rightarrow \infty } \bm F \sim f + g \bm R,
\end{equation}
where $f$ and $g$ are diagonal matrices with elements $f_i$ and $g_i$.
Eq.~\ref{eqn:react} defines the reactance matrix $\bm R$.
The scattering matrix in Eq.~(\ref{eq:S_matrix_def}) is then given by
\begin{align}
\label{SfromR}
  S_{ij} = e^{\ii\xi_i} \, {\left(\frac{1 + \ii \bm R}{1 - \ii \bm R}\right)}_{ij}\!\! \, e^{\ii\xi_j}
\end{align}
for general running indexes $i,j$.
Defined in this way the scattering matrix still has an energy dependence due
to threshold effects.
MQDT gets around this by choosing a new set of reference functions,
${\hat f}_i$, ${\hat g}_i$, defined by WKB-like boundary conditions at
short range, in the classically allowed region of channel $i$
(details of these wave functions are given in Ref.~\citenum{mies:multichannel}).
For our purposes here the key property of these reference functions is that
they are related in a standardized way to the usual energy-normalized asymptotic
reference functions by the transformation
\begin{equation}
\begin{pmatrix} f_i \\ g_i \end{pmatrix} =
\begin{pmatrix} C_i^{-1} & 0 \\ C_i \tan\lambda_i  & C_i \end{pmatrix} \begin{pmatrix} \hat{f}_i \\ \hat{g}_i \end{pmatrix},
\end{equation}
where $C_i(E)$ and $\tan\lambda_i(E)$ are explicitly energy-dependent factors,
with $E$ being the total energy of the system.
The dependence of $C_i$ and $\tan\lambda_i$ with energy has been given explicitly
near threshold~\cite{gao:solutions}, enabling analytical scattering formulas to
be constructed.

The pair of reference functions ${\hat f}_i$ and ${\hat g}_i$ are used as
follows.
Supposing that strong channel couplings lead to a complicated many-channel
scattering wave function, nevertheless there is an intermolecular distance
$r_0$ beyond which the channels are essentially uncoupled (this radius is
indicated schematically in Fig.~\ref{fig:schematic}).
The wave function at radii $r>r_0$ can then be written
\begin{align}
\bm F(r > r_0) \sim {\hat f} + {\hat g}\bm Y,
\label{eq:Y_mat_def}
\end{align}
in terms of a short-range reactance matrix $\bm Y$.
Because of the carefully chosen normalization of ${\hat f}_i$ and ${\hat g}_i$
$\bm Y$ does not carry the energy dependence characteristic of the threshold behavior.
From the standpoint of the threshold $\bm Y$ can be considered constant
(later, we will incorporate an explicit energy dependence due to resonant
states).
The threshold energy dependence is then restored via the transformation
\begin{align}
\bm R = \bm{C}^{-1} {\left[\bm Y^{-1} - \tan \bm \lambda \right]}^{-1} \bm{C}^{-1},
\label{eq:Rmat_from_Ybar}
\end{align}
where $\bm{C}$ and $\tan \bm \lambda$ are the appropriate diagonal
matrices  defined in~\cite{mies:multichannel}.
Alternatively, the energy-independent reference functions can be written in
terms of incoming and outgoing waves
\begin{align}
\label{fplusminus}
  \hat{f}^{\pm}_i =  \hat{g}_i \pm\ii\hat{f}_i.
\end{align}
The scattering wave function can then be represented at short-range by a
scattering matrix $\bar{\bm{S}}$, defined via
\begin{align}
  \bar{\bm{S}} = \frac{1 + \ii\bm Y}{1-\ii \bm Y}
  \label{eq:Sbar_from_Ybar}
\end{align}
with inverse transformation
\begin{align}
  \bm{Y} = \ii \frac{1 -  \bar{\bm S}}{1+  \bar{\bm S}}.
  \label{eq:Ybar_from_Sbar}
\end{align}
The bar notation refers to the ${ S}$-matrix at short-range.
In the next section we will apply the statistical theory approach to highly
resonant scattering from nuclear physics to replace $\bar{\bm{S}}$ with a
suitably energy averaged version, $\braket{\bar{\bm{S}}}$, that is itself energy
independent and includes the absorption effect due to the unobserved and
indirect absorption processes.

\subsection{The short-range $S$-matrix accounting for absorption processes}\label{sec:absproc}
This section details the construction of the short-range energy-averaged
scattering matrix $\bar{\bm S}^\mathrm{abs}$, which accounts for any absorption
processes that may be present in a given experiment.
The elements $\bar{S}^\mathrm{abs}_{ij}$ of this matrix are indexed by the
observable channels $i$, $j =$ $a$, $b$, $c$, $\dots$, $k$, $l$, $m$, $\dots$.
It is constructed so that it may be sub-unitary to account for absorption due
to the two effects described above, direct absorption and complex formation.

The derivation of this matrix proceeds in three steps:
the first constructs an effective, sub-unitary $S$-matrix
$\bar{\bm{S}}^\mathrm{unobs}$ that accounts phenomenologically
for direct absorption to unobserved channels, based on an optical potential;
the second constructs an energy smooth $S$-matrix that accounts for
absorption due to complex formation by averaging a highly resonant $S$-matrix
${\bar{\bm S}}^\mathrm{res}$;
finally, both types of absorption are combined to obtain a matrix
$\bar{\bm S}^\mathrm{abs}$ for the combined absorption processes.
$\bar{\bm S}^\mathrm{abs}$ will then be used in the next section to complete
the  construction of the matrix $\braket{\bar{\bm{S}}}$.

\subsubsection{Absorption due to unobserved channels}\label{sec:unobsloss}
Flux entering in any channel that vanishes due to unobserved processes is
conveniently modeled by incorporating a complex-valued optical potential in
each channel
\begin{equation}
V_i(r) + \frac{\hbar^2 l_i(l_i+1)}{2 m_r r^2} - \ii\frac{\gamma_i(r)}{2},
\label{optpot}
\end{equation}
where $V_i$ is the real-valued potential in the absence of such
absorption for channel $i$~\cite{bethe:continuum,idziaszek.julienne:universal,
osseni.dulieu.ea:optimization}.
$V_i$ therefore incorporates the appropriate long-range interaction
terms for a given system~\cite{gao:general}, and could for example include a
dipole-dipole term~\cite{wang.xie.ea:quantum}.
In the absence of an exact treatment of the short-range interactions using a
full potential energy surface and a full collisional formalism,
the influence of the optical potential $\gamma_i$ is to create a new linear
combination of asymptotic functions.
Specifically,  the short-range $Y$-matrix in Eq.~\eqref{eq:Y_mat_def} can be
replaced in each channel by a purely imaginary quantity
$\ii y_i$~\cite{osseni.dulieu.ea:optimization}.
The wave function in this channel now reads
\begin{equation}\label{eq:ya_def}
  \hat{f}_i + \hat{g}_i \, {\ii y_i} ,
\end{equation}
for a real-valued parameter $y_i$, which we  term the \emph{unobserved
absorption coefficient} in channel $i$.
The optical potential reproduces the overall phenomenological loss from channel $a$
to all the unobserved channels $\rho$, $\sigma$, $\tau$, $\ldots$
Recasting the wave function in terms of incoming and outgoing waves from
Eq.~\eqref{fplusminus} gives
\begin{align}
  \hat{f}_i + \hat{g}_i \, {\ii y{_i}} =& ({\hat f}^+_i + {\hat f}^-_i)\frac{1}{2\ii}
  +({\hat f}^+_i - {\hat f}^-_i)\frac{\ii y{_i}}{2} \nonumber \\
=& \left[{\hat f}^-_i -  f^+_i \left(\frac{1-y_i}{1+y_i}\right) \right]\frac{\ii}{2}(1+y_i).
\end{align}
Here the factor $\frac{\ii}{2}(1+y_i)$ represents an overall normalization,
such that the coefficient of the outgoing wave term in the square brackets gives
the short-range scattering matrix
\begin{equation}
  \bar S_{ij}^\mathrm{unobs} = \left( \frac{1-y_i}{1 + y_i} \right)\delta_{ij}.
\end{equation}
This is a unique overall term for the channel $i$ as the unobserved channels
$\rho$, $\sigma$, $\tau$ $\ldots$ are not explicitly enumerated in
Eq.~\eqref{optpot}.
We note that $\bar{\bm S}^\mathrm{unobs}$ is by definition independent of
energy and as such does not need to be energy-averaged.

The coefficients $y_i$ are purely phenomenological parameters of the
theory.
By considering $0 \le y_i \le 1$, as Ref.~\citenum{idziaszek.julienne:universal} does,
$\bar{\bm S}^\mathrm{unobs}$ is in general sub-unitary, becoming unitary when $y_i=1$.
A special case of this result is in the incident channel, where $i=a$.
In this case, $\lvert \bar{S}_{aa}^\mathrm{unobs}\rvert^2$ is the probability
that the molecules incident in channel $a$ are not lost to the unobserved process.
Then, the (short-range) unobserved absorption probability is given by
\begin{align}
  \bar{p}_\mathrm{unobs} = 1 - {\left(\frac{1-y_a}{1 + y_a}\right)}^2 = \frac{4 y_a}{{(1 + y_a)}^2}.
  \label{eq:loss_probability}
\end{align}

\subsubsection{Absorption due to complex formation}\label{sec:indloss}
The next step is to include the possibility for the scattering wave function to
span the region of the resonances, resulting in indirect absorption due to
resonant complex formation.
The indirect processes all couple to the dense forest of resonant states $\mu$
which results in a highly resonant short-range scattering matrix, here denoted
$\bar{\bm S}^\mathrm{res}$.
As described in Sec.~\ref{sec:highreso}, we average the resonant matrix to get
an energy-smooth scattering matrix $\braket{{\bar{\bm S}^\mathrm{res}}}$.

In order to determine the average of $\bar{\bm S}^\mathrm{res}$ we exploit
the chaotic nature of highly resonant collisions~\cite{croft.bohn:long-lived,
frye.morita.ea:approach,croft.makrides.ea:universality,yang.perez-ros.ea:classical,
croft.balakrishnan.ea:long-lived}, and treat $\bar{\bm S}^\mathrm{res}$
statistically using random-matrix theory (RMT)~\cite{wigner:characteristic,
*wigner:characteristic*1,
dyson:statistical,*dyson:statistical*1,*dyson:statistical*2}.
Here we only sketch the essential steps of the derivation, see
Ref.~\cite{mitchell.richter.ea:random} and references therein for a more
complete treatment.
We first introduce an effective Hamiltonian $H^\mathrm{eff}$ for the resonances,
\begin{equation}
  H^\mathrm{eff}_{\mu \nu} = E_\mu \delta_{\mu \nu} - \ii\pi \sum_i W_{\mu i} W_{i \nu}
  \label{eqn:ham}
\end{equation}
in the diagonal representation, which describes the dynamics of the
resonances~\cite{peskin.reisler.ea:on,mitchell.richter.ea:random}.
Eq.~\eqref{eqn:ham} is based on a partitioning of the Hilbert space into a
bound state space and a scattering channel space, introduced by
Feshbach~\cite{feshbach:optical,feshbach:unified,feshbach:unified*1}.
While Eq.~\eqref{eqn:ham} is complete the effort involved in computing all the
parameters, especially for the large number of bound states that we are
interested in here, make this approach impractical~\cite{feshbach:unified}.
Following RMT, we therefore consider the parameters as purely statistical quantities.
Generally in statistical theories the energies $E_\mu$ of the resonances are
assumed to form a distribution whose nearest-neighbor spacing statistics satisfy
the Wigner-Dyson distribution with mean level spacing $d$ and the coupling
matrix elements $W_{i \mu}$, between a channel $i$ and a resonant state $\mu$,
are assumed to be Gaussian random variables with vanishing mean and second moment
\begin{equation}
  \braket{W_{\mu i} W_{\nu j}} = \delta_{\mu \nu}\delta_{i j} \, \nu_i^2,
\end{equation}
where $\nu_i$ is the magnitude of the bound state-scattering channel coupling
for channel $i$.
However we will not need to specify the distributions for our present purposes.

In terms of the MQDT reference functions $\hat{f}_i$, $\hat{g}_i$ defined above
resonant scattering will result in a short-range ${\bm Y}^\mathrm{res}$ matrix
\begin{equation}
  \bm F \sim \hat{f} +  \ \hat{g} \bm Y^\mathrm{res},
\label{eq:K_sr_def}
\end{equation}
similar to Eq.~\eqref{eq:Y_mat_def} with the running indices
$i,j = a$, $b$, $c$, $\ldots$, $k$, $l$, $m$, $\ldots$
It is assumed that all the resonant states have outer turning points at
distances $r<r_0$, so that ${\bm Y}^\mathrm{res}$ contains the full structure
of the resonances and therefore takes the form~\cite{mitchell.richter.ea:random}
\begin{align}
  Y_{ij}^\mathrm{res} = \pi \sum_{\mu} \frac{W_{i \mu} W_{\mu j}}{E - E_{\mu}}.
  \label{eqn:Yres_def}
\end{align}
In the weak-coupling limit, $\nu_i \ll d$ the resonances are isolated and can,
if desired, be described in terms of resonant widths given by
$\gamma_{\mu} = 2\pi\sum_i |W_{i\mu}|^2$.

In order to average over many resonances, we use the statistical independence
of the coupling matrix elements,
\begin{align}
\braket{W_{i \mu} W_{\mu a}}  = \delta_{ia} \nu_i^2,
\end{align}
to assert that $\braket{{\bm Y}^\mathrm{res}}$ is diagonal.
This implies that $\braket{\bar{S}_{ij}^\mathrm{res}} = 0$ if $i \ne j$ which
confirms that ${\bm Y}^\mathrm{res}$ contains no direct contribution to the
scattering and describes purely resonant scattering as desired.
Moreover, the average of ${\bm Y}^\mathrm{res}$ over many isolated resonances
is equivalent to the average over a single representative resonance.
As the resonances are separated on average by a spacing energy $d$ this average
becomes
\begin{align}
\braket{Y_{ii}^\mathrm{res}} = \frac{1}{d} \int_{E_{\mu} - d/2}^{E_{\mu}+d/2} \!\!\! dE \, \frac{\pi\nu_i^2}{E-E_{\mu}}.
\end{align}
Evaluating this integral in the principal value sense gives
\begin{align}
  \braket{Y_{ii}^\mathrm{res}} & \approx \frac{\pi \nu_i^2}{d} \lim_{t \rightarrow 0^+} \mathcal{P} \int_{-\infty}^{\infty} \!\!\! \diff E \, \frac{\exp(\ii t E)}{E} \nonumber \\
                               & = \ii \frac{\pi^2 \nu_i^2}{d},
\end{align}
where the contour is closed on the upper half of the complex plane.
Therefore, in each channel, the wave function accounting for indirect absorption
due to complex formation is given by
\begin{align}
  \hat{f}_i +  \hat{g}_i \, {\ii x_i},
  \label{eqn:funcformx}
\end{align}
in terms of a real-valued parameter
\begin{align}
x_i = \frac{\pi^2 \nu_i^2}{d},
\label{eqn:x}
\end{align}
which we term the \emph{indirect absorption coefficient} in channel $i$.
Although coming from an entirely different mechanism, the form of the
wave function in Eq.~\eqref{eqn:funcformx} is exactly the same as that for
unobserved absorption given by Eq.~\eqref{eq:ya_def}.
Similarly, this leads to a short-range scattering matrix in each channel $i$ that
corresponds to indirect absorption
\begin{align}
  \braket{\bar{S}_{ij}^\mathrm{res}}= \left(\frac{1-x_i}{1 + x_i}\right)\delta_{ij}.
\label{eq:average_resonant_S}
\end{align}
This is a unique term for the channel $i$ due to the fact that the off-diagonal
elements are zero.
Again specializing to the case of the incident channel, $i=a$, the
(short-range) indirect absorption probability in channel $a$ is given by
\begin{equation}
  \bar{p}_\mathrm{res} = 1 - {\left(\frac{1-x_a}{1 + x_a}\right)}^2 = \frac{4 x_a}{{(1 + x_a)}^2}.
  \label{eq:loss_probability_res}
\end{equation}

The unobserved and indirect absorption cases are formally similar.
In both cases, the short-range scattering matrix is given by an absorption
coefficient $y_i$ and $x_i$ in each channel $i$.
For unobserved processes the absorption coefficient $y_i$ is a purely
phenomenological fitting parameter, whereas for indirect processes the indirect
absorption coefficient $x_i$ contains information about the complex itself, namely,
the ratio of bound states-scattering channel coupling to the mean level spacing.
The form of $x_i$ is reminiscent of Fermi's golden rule as it connects the
average scattering matrix to the square of the bound-continuum matrix element $\nu_i^2$
and the density of states $\rho = 1/d$.
It should be noted that this result is quite general, and need not rely on the
assumption of weak coupling.
Eq.~\eqref{eqn:x} can be derived in a number of ways: using a Born expansion of
the $S$-matrix~\cite{agassi.weidenmuller.ea:statistical,
mitchell.richter.ea:random};
via the replica trick~\cite{weidenmuller:statistical};
or using the supersymmetry approach~\cite{efetov:supersymmetry,
verbaarschot.zirnbauer:critique}.

\subsubsection{Total absorption}\label{sec:absloss}
\begin{figure}[tb]
\centering
\includegraphics[width=0.99\columnwidth]{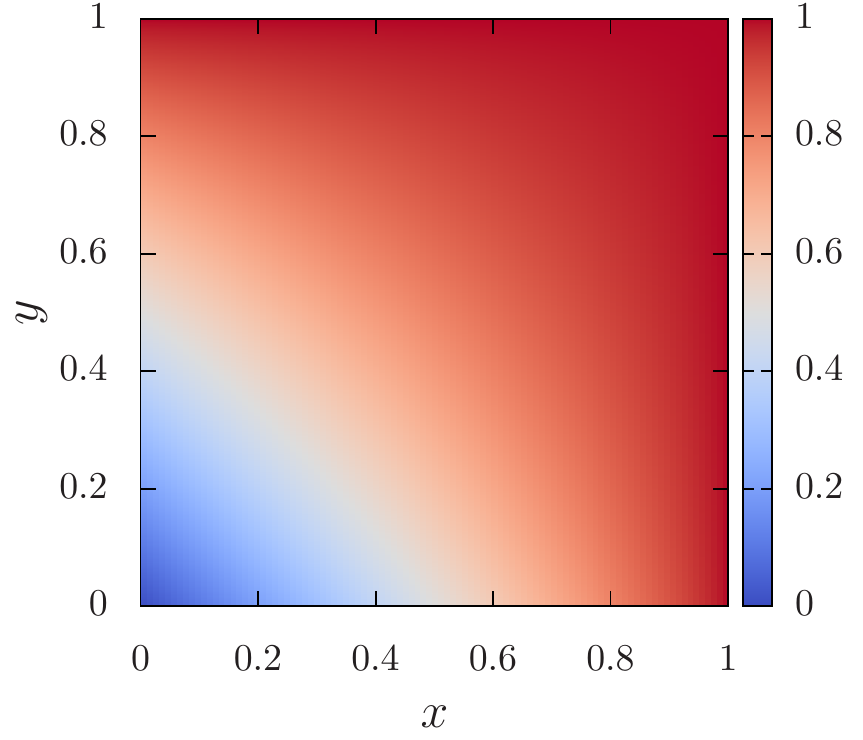}
\caption{Interplay of the dimensionless indirect and unobserved absorption
coefficients $x$ and $y$ on the dimensionless absorption coefficient $z$ given
by Eq.~\eqref{eq:zformula}.
}\label{fig:z}
\end{figure}

As the unobserved and indirect cases are formally similar, we can combine both
processes to recover the total absorption process in Eq.~\eqref{XS-tilde}.
The resulting short-range scattering matrix  is given by
\begin{align}
\label{eqn:compoloss}
  \bm{\bar S}^\mathrm{abs} = \bm{\bar S}^\mathrm{unobs} \braket{{\bm{\bar S}^\mathrm{res}}}
\end{align}
so that
\begin{align}
  \bar S_{ij}^\mathrm{abs}
     =  \left(\frac{1 - y_i}{1 + y_i}\right) \left(\frac{1-x_i}{1+x_i}\right)\delta_{ij}.
\end{align}
The two kinds of effects can be consolidated into a unified form
\begin{align}
  \bar S_{ij}^\mathrm{abs} = \left(\frac{1-z_i}{1 + z_i} \right)\delta_{ij} \label{eq:Swithz}
\end{align}
in terms of an effective \emph{absorption coefficient}
\begin{align}
  z_i = \frac{x_i + y_i}{1 + x_i y_i}
\label{eq:zformula}
\end{align}
that combines both unobserved and indirect types of absorption.
Fig.~\ref{fig:z} illustrates Eq.~\eqref{eq:zformula} and the interplay of the
different values $x_i$ and $y_i$.  Notice that if either $y_i$ or $x_i$ should be zero, then $z_i$ automatically
reverts to the other one.

Once again, specializing to the case of the incident channel $i=a$,
The (short-range) absorption probability in the incident channel $a$ as defined in Eq.~\eqref{probaabs} is then given by
\begin{align}
  \bar{p}_\mathrm{abs} = 1 - {\left(\frac{1-z_a}{1 + z_a}\right)}^2 = \frac{4 z_a}{{(1 + z_a)}^2}.
  \label{eq:loss_probability_tot}
\end{align}
Therefore, even in the presence of both types of absorption, if the only
observable fact is that absorption has occurred, then a value $z_a$
(or an equivalent parametrisation) can be extracted, as has been done in several
studies~\cite{kotochigova:dispersion,idziaszek.julienne:universal,
idziaszek.quemener.ea:simple,gregory.frye.ea:sticky,frye.julienne.ea:cold,
bai.li.ea:model,li.li.ea:universal,bai.li.ea:simple}.

Finally, the coefficients $z$, $y$ or $x$ can also depend implicitly on
different experimental tools of control such as an electric field
$\mathcal{E}$~\cite{guo.ye.ea:dipolar}; a magnetic field $B$;
or the intensity of surrounding electromagnetic waves, whether it is due to the
surrounding trapping laser~\cite{christianen.zwierlein.ea:photoinduced},
red-detuned photo-association~\cite{perez-ros.lepers.ea:theory},
or blue-detuned shielding~\cite{lassabliere.quemener:controlling,
karman.hutson:microwave,karman:microwave}.
Therefore, the coefficients should carry such dependence so that
$z = z(\mathcal{E}, B, I)$,  similarly for $x$ and $y$.
We omit this dependence to simplify the notations in the following, unless stated otherwise.

\subsection{The complete short-range $S$-matrix}\label{complete}
We now combine the short-range absorption scattering matrix
$\bm{\bar S}^\mathrm{abs}$, which gathers the unobserved direct processes and
all the indirect processes, with a short-range direct scattering matrix
$\bm{\bar S}^0$ that includes all the direct processes (elastic, inelastic,
reactive) to obtain $\braket{\bm{\bar{S}}}$---which is the energy average of
the physical short-range $S$-matrix of the system of interest $\bm{\bar{S}}$.

The starting point is a short-range unitary scattering matrix $\bm{\bar S}^0$
that contains all the direct processes such as the molecular elastic, inelastic
and reactive scattering in the absence of any absorption due to unobserved
channels or complex formation.
For example $\bm{\bar S}^0$ could be obtained from a scattering calculation
containing only asymptotically open channels.
As $\bm{\bar S}^0$ contains no resonances it is by definition energy insensitive
and there is no need to take its energy average.
$\bm{\bar S}^0$ is defined at $r_0$ by a wave function of the form
\begin{equation}\label{generalform}
  \hat{f}^- - \hat{f}^+ \bm{\bar S}^0
\end{equation}
restricting, as usual, the indices $i,j$ of the matrices to the observed channels
$i,j=$ $a$, $b$, $c$, \ldots, $k$, $l$, $m$, \ldots $\,$
Starting with this foundation, we transform $\bm{\bar S}^0$ to include the
effect of the absorption processes contained in $\bm{\bar S}^\mathrm{abs}$.

Generically, a short-range process, 1, makes itself apparent through the linear
combination of outgoing waves that accompany a certain flux of incoming waves,
thus process 1 defines the long-range wave function at $r>r_0$,
\begin{align}
  \bm F_1 = \left[ f^- - f^+ \bm S_1 \right]\bm{\mathcal{N}} ,
\end{align}
where $f^-$ and $f^+$ are diagonal matrices consisting of incoming and
outgoing channel wave functions, respectively, and $\bm{\mathcal{N}}$ is a
overall normalization matrix that is not relevant to our purposes here.
We are interested in how the linear combination of $f^-$ and $f^+$ would change
if a second short-range process, for example absorption, would be introduced.

To this end, we first diagonalize $\bm S_1$ so as to write it in terms of its
eigenphases $\delta_{\alpha}$,
\begin{align}
  {(S_1)}_{ij} = \sum_{\alpha} \braket{i|\alpha} \exp(2\ii \delta_{\alpha}) \braket{\alpha|j}.
\end{align}
It is then possible to define the square root of this matrix,
\begin{align}
  {(S_1^{1/2})}_{ij} &= \sum_{\alpha} \braket{i|\alpha}\exp(\ii \delta_{\alpha}) \braket{\alpha|j}.
\end{align}
The wave function in the presence of process $1$ can then be written
\begin{align}
  \bm F_1 = \left[f^-\bm S_1^{-1/2}  - f^+ \bm S_1^{1/2}  \right]\bm S_1^{1/2}\bm{\mathcal{N}},
\end{align}
thus effectively defining a new set of incoming and outgoing reference
functions $f^- \bm S_1^{-1/2}$ and $f^+ \bm S_1^{1/2}$.
We now introduce a second short-range scattering process, which changes the
boundary condition of the wave function at the asymptotic matching radius $r_0$.
This new scattering wave function, due to both processes, can be written as a
linear combination of the new incoming and outgoing waves defining a second
scattering matrix $\bm S_2$
 \begin{align}
   \bm F_{1,2} = \left[(f^- \bm S_1^{-1/2}) - (f^+ \bm S_1^{1/2})\bm S_2  \right],
 \end{align}
where we have left off another arbitrary normalization.
Finally, factoring out $\bm{S_1}^{-1/2}$ gives
\begin{align}
  \bm F_{1,2} = \left[f^- - f^+ \bm S_1^{1/2} \bm S_2 \bm S_1^{1/2}\right]\bm S_1^{-1/2}.
\end{align}
This identifies the joint scattering matrix for both processes together as
\begin{align}
  \bm S_{1,2} = \bm S_1^{1/2} \bm S_2 \bm S_1^{1/2}.
 \label{eqn:s_composition}
\end{align}
Notice that when we shut off process 1 by setting $\bm S _1 = \bm I$ we get
$\bm S_{1,2} = \bm S_2$, the scattering matrix in the absence of process 1.
Consider for example a single-channel, with $S_1 = \exp(2\ii \delta_1)$,
$S_2 = \exp(2\ii \delta_2)$ then Eq.~(\ref{eqn:s_composition}) simply asserts
that the phase shifts add.
This approach to combine $S$-matrices can be applied recursively to include
other processes as desired.

Using Eq.~\eqref{eqn:s_composition} can combine $\bm{\bar S}^0$ and
$\bm{\bar{S}}^\mathrm{abs}$
\begin{align}
  \braket{\bm{\bar{S}}} = {\left(\bm{\bar S}^0 \right)}^{1/2} \bm{\bar{S}}^\mathrm{abs} {\left(\bm{\bar S}^0 \right)}^{1/2}.
  \label{eqn:full_s}
\end{align}
Eq.~\eqref{eqn:full_s} is the main result of the paper and is quite general.
Of course from Eq.~\eqref{eqn:compoloss}, $\bm{\bar{S}}^\mathrm{abs}$ in
Eq.~\eqref{eqn:full_s} can reduce to either $\bm{\bar{S}}^\mathrm{unobs}$ or
$\braket{\bm{\bar{S}}^\mathrm{res}}$ if only unobserved or indirect absorption
is present.
Therefore, the size of the matrix considered in Eq.~\eqref{eqn:full_s} can be
as high as the number of elastic, inelastic and reactive channels that are
observed in an experiment being modeled.

\section{The two-channel case}\label{sec:2chan}
\begin{figure}[tb]
\centering
\includegraphics[width=0.99\columnwidth]{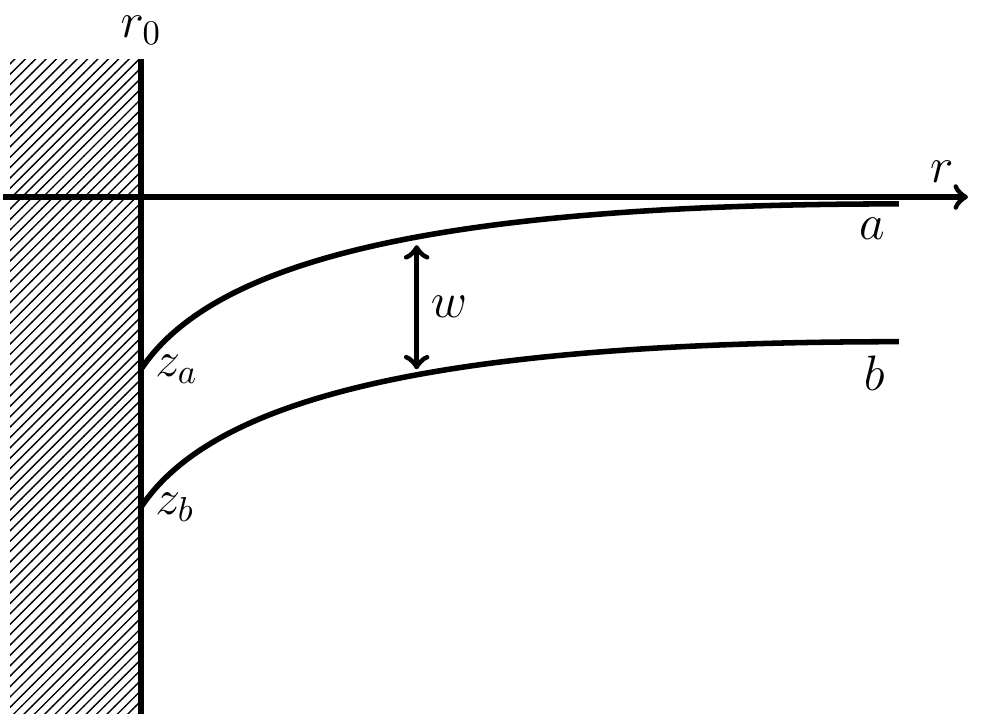}
\caption{Schematic figure showing the two-channel case.
Potential energy curves are plotted versus intermolecular distance $r$
(both in  arbitrary units).
The two channels are labelled $a$ and $b$, while the coupling between the
channels is characterized by the dimensionless quantity $w$.
At short-range, absorption from direct and/or indirect processes is described by
an effective dimensionless absorption coefficient $z_a$, $z_b$ in each channel.
}\label{fig:schematic_closeup}
\end{figure}

Rather than study any particular system, in order to gain insight we consider
the case with just two open channels.
Consisting of an incident channel $a$ and one additional observed channel $b$,
with channel $a$ higher in energy as illustrated in Fig.~\ref{fig:schematic_closeup}.
We emphasize that while we label the second channel with a $b$, which identifies
the channel with an inelastic scattering process, it could equally well be
labelled $k$, and be identified with a reactive scattering process.
Whatever the nature of the scattering to channel $b$ is, we contemplate the
scattering process $a \to b$ and the influence that absorption has on this
process.

Upon reaching short range $r=r_0$, each channel experiences some kind of
absorption with coefficient $z_a$, $z_b$, whose exact origin we do not worry
about here.
It could consist of direct scattering to unobserved channels, or to complex
formation, or some combination of both.
In the absence of these processes the two channels would be somehow coupled and
scattering from one to the other could occur when they are both open.
Using the convention in~\cite{idziaszek.julienne:universal,
jachymski.krych.ea:quantum-defect},
we select reference functions $\hat{f}$ and $\hat{g}$ in each channel so that
the diagonal matrix elements of the short-range matrix $\bm Y$
vanish~\cite{giusti-suzor.fano:alternative},
as such the scattering length in each channel may appear in the MQDT parameters
$C$ and $\tan \lambda$ introduced in Sec.~\ref{sec:mqdt}.
The short-range matrix ${\bm Y}^0$ is therefore defined by a single, real-valued,
off-diagonal coefficient $w$ via
\begin{equation}
{\bm Y}^0 = \begin{pmatrix} 0 & \sqrt{w} \\ \sqrt{w} & 0 \end{pmatrix}.
\end{equation}
This notation uses the letter $w$ to designate the coupling between the observed
channels, since $y$ was already used above to denote the unobserved absorption
coefficient.
This matrix $\bm{Y}^0$ gives the form of the unitary short-range matrix
$\bm{\bar S}^0$ that characterises the direct inelastic/reactive scattering as
\begin{equation}
  \bm{\bar S}^0 = \frac{1 + \ii \bm{Y}^0 }{1 - \ii\bm{Y}^0} = \frac{1}{1 + w}
\begin{pmatrix} 1-w & 2\ii\sqrt{w} \\ 2\ii\sqrt{w} & 1-w \end{pmatrix}
\end{equation}
as also given in Ref.~\cite{jachymski.krych.ea:quantum-defect}.
Written in this way, we could of course regard scattering from $a$ to $b$ as
yet another absorption process, writing $\bar{S}_{aa}^0 = (1-w)/(1+w)$ and identifying
an inelastic/reactive absorption coefficient $w$.
However here we are interested in the prospect of observing the molecular product
directly, and so to treat the scattering matrix element $\bar{S}_{ab}^0$ explicitly.
From Eq.~\eqref{eq:Swithz}, we have
\begin{equation}
\bm{\bar S}^\mathrm{abs} = \begin{pmatrix} \frac{1-z_a}{1+z_a} & 0 \\ 0 & \frac{1-z_b}{1+z_b} \end{pmatrix}.
\end{equation}
To construct the complete short-range matrix $\braket{\bm{\bar{S}}}$ using
Eq.~\eqref{eqn:full_s}, we require the square root of $\bm{\bar S}^0$,
which is given by
\begin{equation}
  {\left( \bm{\bar S}^0 \right)}^{1/2} = \frac{1}{\sqrt{1 + w}} \begin{pmatrix} 1 & \ii\sqrt{w} \\ \ii\sqrt{w} & 1 \end{pmatrix},
\end{equation}
from which we obtain
\begin{align}
\label{eq:two_chan_Smat_full}
  \braket{\bm{\bar{S}}} &= {\left( \bm{\bar  S}^0 \right)}^{1/2} \bm{\bar{S}}^\mathrm{abs} {\left(\bm{\bar S}^0 \right)}^{1/2}  \\
&=  \frac{1}{1+w}  \begin{pmatrix} r_a-w \, r_b & \ii\sqrt{w}(r_a+r_b) \\ \ii\sqrt{w}(r_a + r_b) & r_b - w \, r_a \end{pmatrix}, \nonumber
\end{align}
using the shorthand notation $r_i = (1-z_i)/(1+z_i)$, $i=a$,$ b$.
The expression in Eq.~\eqref{eq:two_chan_Smat_full} is quite general.
It is however instructive to make the assumption that $z_a=z_b=z$, that is
the two channels experience the same absorption, to simplify results and
gain intuition.
In this case the short-range matrix $\braket{\bm{\bar{S}}}$ simplifies to
\begin{equation}
\braket{\bm{\bar{S}}} = \frac{1}{(1+w)(1+z)} \begin{pmatrix} (1-w)(1-z) & 2\ii\sqrt{w}(1-z) \\ 2\ii\sqrt{w} (1-z) & (1-w)(1-z) \end{pmatrix}.
\label{eq:two_chan_Smat}
\end{equation}
Notice that, in a case where channel $b$ were \emph{not} observed, this model
would return to the single matrix element
$\braket{\bm{\bar{S}}}_{aa} = (1-q)/(1+q)$, written in terms of an effective
absorption coefficient
\begin{equation}
q = \frac{w+z}{1+wz}
\label{eq:q_def}
\end{equation}
that describes composite absorption from the combination of inelastic/reactive
scattering to unobserved channel with absorption due to complex formation.
This nicely illustrates the flexibility of the model to treat channels as
either observed or unobserved, as required.

\begin{figure}[tb]
\centering
\includegraphics[width=0.71\columnwidth]{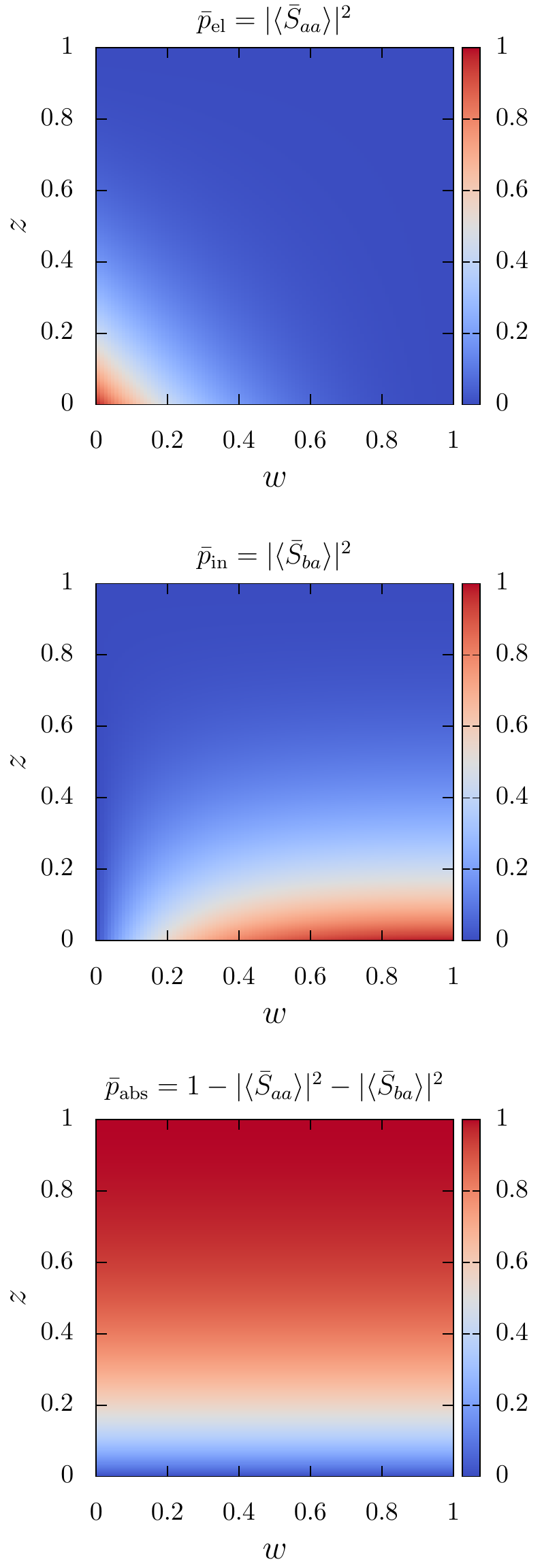}
\caption{Top to bottom: short-range probabilities for elastic scattering,
inelastic scattering, and absorption, versus the
dimensionless quantity $w$ which characterizes the inter-channel coupling
and the dimensionless absorption coefficient $z$.
}\label{fig:proba2lev}
\end{figure}

To see the basic interplay between direct scattering and absorption,
it is worthwhile to consider the probabilities for various outcomes, shorn of
the additional complications of threshold effects.
Eq.~\eqref{eq:two_chan_Smat} encodes three types of probability:
\begin{itemize}
\item the elastic scattering probability due to a direct process for the
  incident channel $a$,
\begin{equation}
\label{probasrqu}
  \bar{p}_\mathrm{el} = \lvert\braket{{\bar S}_{aa}}\rvert^2 = \frac{{(1-w)}^2{(1-z)}^2}{{(1+w)}^2{(1+z)}^2}.
\end{equation}
\item the probability to enter in channel $a$ and emerge in channel $b$ due to
a direct process,
\begin{equation}
\label{probasrin}
 \bar{p}_\mathrm{in}  = \lvert\braket{{\bar S}_{ba}}\rvert^2 = \frac{4w}{{(1+w)}^2} \frac{{(1-z)}^2}{{(1+z)}^2}.
\end{equation}
\item the absorption probability due to unobserved channels and/or complex formation
\begin{align}
\label{eq:two_chan_abs}
\bar{p}_\mathrm{abs}  &= 1- \lvert\braket{{\bar S}_{aa}}\rvert^2 - \lvert\braket{{\bar S}_{ba}}\rvert^2 \nonumber \\
                        &=\frac{4z}{{(1 + z)}^2},
\end{align}
where we recover Eq.~\eqref{eq:loss_probability_tot}.
\end{itemize}
These various probabilities are shown in Fig.~\ref{fig:proba2lev} as a function
of the interchannel coupling $w$ and the absorption coefficient $z$.
As seen in the top figure, the elastic probability is quite low unless both
$w$ or $z$ have a low value.
In the other panels the influence of each process on the other can be
appreciated.
The middle panel shows that direct scattering only happens with appreciable
probability when the absorption coefficient $z$ is small, below around 0.2.
On the other hand, the absorption probability is indifferent to coupling strength
$w$ as can be seen on the bottom panel and implied by
Eq.~\eqref{eq:two_chan_abs} which is independent of $w$.
This kind of absorption is a one-way journey: incident flux that gets to short range
is lost and will not emerge from channel $b$.
Note that the coefficients $w$ and $z$ are not interchangeable and do not play
equivalent roles in the theory.
One observable actually pertains to seeing the scattering end in a particular
observable channel, whereas the other is simply absorption.

So far, we have just dealt with the short-range scattering matrix
$\braket{\bm{\bar{S}}}$.
In order to get the asymptotic $S$-matrix, $\braket{\bm{S}}$, we need to
include threshold effects using MQDT\@.
The QDT parameters $C_a$ and $\tan \lambda_a$ in Eq.~\eqref{eq:Rmat_from_Ybar}
are known analytically for s-wave threshold collisions for a $1/r^6$
long-range potential~\cite{gao:solutions,idziaszek.julienne:universal}
\begin{align}
C_a^{-2} &\approx k \bar{\mathrm{a}} (1+{(s_a- 1)}^2), \nonumber \\
\tan \lambda_a &\approx 1-s_a,
\label{eq:approx_QDT_parameters}
\end{align}
where $s_a = \mathrm{a}/\bar{\mathrm{a}}$, $\mathrm{a}$ being the scattering
length in channel $a$, $\bar{\mathrm{a}} = 2 \pi  R_6 / \Gamma (1/4) $ is the Gribakin-Flambaum
length~\cite{gribakin.flambaum:calculation,jachymski.krych.ea:quantum-defect},
$R_6 = {(2 \mu C_6 / \hbar^2)}^{1/4}$ is the van der Waals length.
We further assume that  channel $b$ is far from threshold, as such
$C_{b}=1$, $\tan \lambda_{b}=0$.
Following the approach outlined in Section~\ref{sec:mqdt}, $\braket{\bm{S}}$
and the corresponding cross sections and rate coefficients, can be obtained from
$\braket{\bm{\bar{S}}}$ and the QDT parameters in each channel.

We first consider the case, where no exit channel is explicitly observed.
In this case the quenching coefficient $q$ in (\ref{eq:q_def}) plays the role
of an absorption coefficient.
Then $q$ can directly replace $y$ in the quenching formulas of
Ref.~\cite{jachymski.krych.ea:quantum-defect}. For example, the physical quenching probability is
\begin{align}
p_\mathrm{qu} \simeq \bar{p}_\mathrm{qu} \times \frac{C^{-2}(1+q)^2}{(1 + q \, C^{-2})^2 + q^{2} \, \tan^2 \lambda } .
\label{probalrqu}
\end{align}
Here the factor $\bar{p}_\mathrm{qu} = 4q/(1+q)^2$ is the probability for quenching given
that the molecules get close together,
while the final factor modifies this probability due to the quantum reflection effects
that modify the molecules chances of getting close together.
This effect has been discussed at length elsewhere \cite{ jachymski.krych.ea:quantum-defect} and we do not repeat the
discussion here.

In terms of the elastic and quenching rate coefficient, we have
from Eq.~(26) and Eq.~(28) of Ref.~\cite{jachymski.krych.ea:quantum-defect}
\begin{align}
  \tilde{\beta}_\mathrm{el} &= g\frac{4\pi\hbar}{\mu} k \bar {\text{a}}^2 \frac{s_a^2 + {\left(\frac{w+z}{1+wz} \right)}^2 {(2 - s_a)}^2}{1 + {\left(\frac{w+z}{1+wz} \right)}^2 {(s_a-1)}^2} \nonumber \\
   \tilde{\beta}_\mathrm{qu} &= g\frac{4\pi\hbar}{\mu} \bar {\text{a}} \left(\frac{w+z}{1+wz} \right) \frac{1 + (s_a - 1)^2}{1 +\left(\frac{w+z}{1+wz} \right)^2 (s_a - 1)^2} .
   \label{eqn:ratemqdt}
\end{align}
Of course when $z=0$, two channels with inelastic collisions but no absorption,
$\tilde{\beta}_\mathrm{qu}$ identifies with
\begin{equation}
  \tilde{\beta}_\mathrm{in} = g\frac{4\pi\hbar}{\mu}\bar{\text{a}}w\frac{1 + {(s_a - 1)}^2}{1 + w^2{(s_a - 1)}^2}
\end{equation}
or when $w=0$, one channel with absorption but no coupling to inelastic channels,
$\tilde{\beta}_\mathrm{qu}$ identifies with
\begin{equation}
  \tilde{\beta}_\mathrm{abs}= g \, \frac{4 \pi  \hbar}{\mu} \, \bar {\text{a}}  \, z \, \frac{1 + (s_a - 1)^2}{1 + z^2 \, (s_a - 1)^2}
\end{equation}
which are the equations found previously in
Ref.~\cite{jachymski.krych.ea:quantum-defect}.

\begin{figure}[tb]
\centering
\includegraphics[width=0.99\columnwidth]{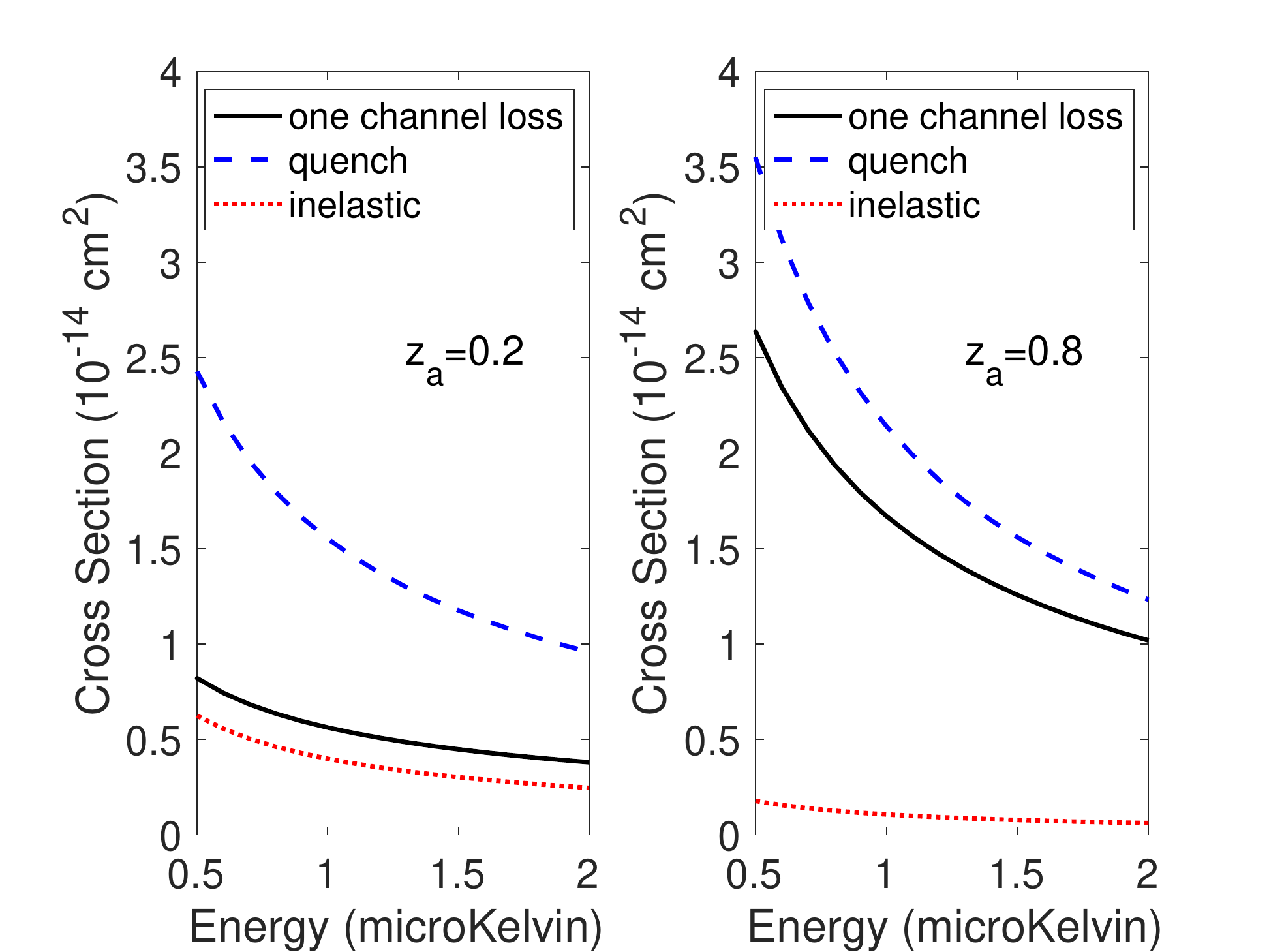}
\caption{Schematic cross sections for NaRb molecules colliding in an excited
  state for two different values of $z_a =$ 0.2 (left panel) and 0.8 (right panel).
  Shown for the two-channel case are the quenching (blue dashed) and inelastic
  (red dotted) cross sections as computed from Eqs.~\eqref{probasrqu},\eqref{probasrin},
  and~\eqref{eq:two_chan_Smat}, fixing the dimensionless quantities $z_a=0.5$,
  $s_a=1.0$, and $w=1.0$.
  For comparison, the black curve shows the cross section for the single-channel
  case ($w=0$) with an absorption coefficient $z_a$ and phase factor $s_a$.
}\label{fig:probaNaRb}
\end{figure}

As a simple illustration of the relation between direct scattering and absorption,
Fig~\ref{fig:probaNaRb} shows several representative cross sections for the
two-channel case.
For concreteness, we show cross sections for molecules with the mass and $C_6$
coefficient of NaRb.
In this case, there are several undetermined coefficients, $z_a$, $z_b$,
$s_a$, $s_b$, and $w$, likely too many to make a meaningful fit to the NaRb data.
For this illustration we have somewhat arbitrarily set $z_b=0.5$, set the incident
channel scattering length to $s_a = 1.0$, and set the interchannel coupling to
$w=1.0$, which would give the maximum inelastic scattering in the absence of
absorption.  Note that the phase parameter $s_b$ in the final channel is
irrelevant, as this channel is assumed far from threshold.

The left and right panels in the figure give results for absorption coefficients
in the incident channel of $z_a=0.2$ and $z_a=0.8$, respectively.
In each panel, results from an explicitly two-channel model are shown in color.
Specifically, the blue and red curves describe the quenching and inelastic
cross sections, respectively.
Unsurprisingly, the total quench cross section is greater than the cross
section for inelastic scattering alone.
As a comparison, the black line shows the cross section that results if we use
a one-channel scattering model with the same absorption coefficient $z=z_a$ and
phase parameter $s=s_a$ in that channel.
It is seen that the inelastic process alters the quenching cross section
significantly.
The right panel repeats this calculation, for a larger incident channel
absorption coefficient $z_a=0.8$.
This larger value of $z_a$ both raises the total quenching cross section, and reduces
the relative cross section for inelastic scattering.

In practice, if both the quenching and inelastic cross sections were measured,
their energy-dependent cross sections (or temperature-dependent rate coefficients)
could be simultaneously fit by formulas such as these, yielding consistent values
of the absorption coefficients $w$, $z_a$, and $z_b$, and the incident phase parameter $s_a$.

\section{Application to various experimental situations}
Cast in terms of the present theory it is interesting to draw some tentative
conclusions about the experiments that have been performed so far.
Quantitative description will likely require further information, yet the basic
formulas Eq.~\eqref{eq:loss_probability_tot} and Eq.~\eqref{eq:zformula}
may guide our thinking.
Note that in this section we remove the subscript $a$ of the absorption coefficients,
for clarity.

\subsection{Collisions of endothermic processes}
The most basic collision of ultracold molecules is one in which both molecules
are in their absolute ground state and are not chemically reactive at a
temperature low enough such that all other channels are asymptotically closed.
This was achieved in collisions of NaRb molecules~\cite{ye.guo.ea:collisions}
and RbCs molecules~\cite{gregory.frye.ea:sticky}.
In this case the presumed losses are due to complex formation, as such $z$
reduces to $x$.
Light is present in these experiments and may strongly affect the molecular
losses~\cite{gregory.blackmore.ea:loss}, therefore $x$ should in general depend
on $I$, the intensity of the trapping laser.

Knowing the energy dependence of the loss cross section enables one to extract
the absorption coefficients from experimental data.
Thus for NaRb collisions, the fitted parameter gave
$x(I) = 0.5$~\cite{bai.li.ea:model} while for RbCs collisions it gave
$x(I) = 0.26(3)$~\cite{gregory.frye.ea:sticky}, at the specific laser light
intensities of these experiments.
While the intensity dependence of $x$ remains unknown, we can make the
following assumption.
If the light absorption is saturated with intensity we assume that any complex
that is formed decays immediately.
In this case the value of $x$ is a measure of the formation of the complex,
and represents a direct measure of the ratio between the mean bound
state-scattering channel coupling and the mean level spacing of resonances in
the complex at the specific laser light intensities.
Interpreting the mean coupling as a mean width,
via $\nu^2 = \gamma/2 \pi$, Eq.~(\ref{eqn:x}), would give the ratio of mean
resonance width $\gamma$ to mean level spacing $d$, $\gamma/d = 2 x / \pi $.
The ratio would be $\gamma/d = 0.32$ for NaRb and $\gamma/d = 0.17$ for RbCs.
Thus the theory produces from the data a concrete prediction that can be used
to test a microscopic theory of molecular collisions.
This interpretation relies on the assumption that the measured $x$ in the
presence of the light truly represents the complex formation.
This assumption would of course not be necessary if the measurement were
repeated in a box trap.

Even under this assumption, the comparison between empirical and calculated
values of $\gamma/d$ is complicated by the presence of external fields in the
experiment, as $x$ can also depend on those fields in addition to the intensity,
so that $x = x(\mathcal{E},B,I)$.
For a pure field-free case (no electric field nor magnetic field),
a coupled representation scheme can be used to estimate $d$ when the total
angular momentum quantum number $J$ and its laboratory projection $M$ are
conserved~\cite{christianen.karman.ea:quasiclassical}.
However, even though the electric field is zero, the NaRb and RbCs measurements
are performed in a non-zero magnetic field.
It seems therefore appropriate to include in the microscopic estimates,
collections of states with different values of $J$ that are mixed by the field.
It remains uncertain, however, how many values of $J$ are relevant to the
estimate of $\gamma/d$ for a given magnetic field value.
It should also be noted that the application of an electric field appears to
alter the absorption coefficient, raising it to the universal value
$x(\mathcal{E}>0)=1$~\cite{guo.ye.ea:dipolar},
indicating that the electric field increases the strength of channel coupling,
the density of resonant states, or perhaps both.

\subsection{Chemically Reactive Collisions}
An alternative set of experiments, spanning the past decade, has measured loss
in ultracold KRb molecules, distinguished from NaRb or RbCs in that the
KRb + KRb $\to$ K$_2$ + Rb$_2$ reaction is exothermic.
In the pioneering experiments~\cite{ospelkaus.ni.ea:quantum-state},
the products were not observed, therefore reactive scattering contributed to
the unobserved processes described by the coefficient $y$ in addition to the
$x$ in the endothermic case.
In general, these experiments exhibit loss consistent with an absorption
coefficient $z(I)=1$ from Eq.~\eqref{eq:zformula},
corresponding to loss of all molecules that get close enough to react or form
a collision complex~\cite{de-marco.valtolina.ea:degenerate}.
However, Eq.~\eqref{eq:zformula} does not lead to the identification of the
separate mechanism for unobserved (chemical reaction in that case) and
indirect (complex formation) processes.
Nevertheless, the existence of both processes has been verified experimentally,
by the identification in REMPI spectroscopy of both the products K$_2$ and Rb$_2$,
and the intermediaries K$_2$Rb$_2^+$~\cite{hu.liu.ea:direct}.
From Eq.~\eqref{eq:zformula}, we note that in this parametrization $z(I)=1$ can
occur only if $x(I)=1$ or $y(I)=1$.
It seems likely that indirect loss from complex formation does not occur with
unit probability, that is, $x(I)$ is likely less than unity, since this is
certainly the case for the non-reactive species NaRb and RbCs (see above).
The difference in energies that renders the KRb reaction exothermic, a mere
10~\si{\wn}, is decided at long range as the products recede from one another,
and likely has little bearing on the complex itself and the couplings that
determine the value of $x(I)$.
We therefore provisionally conclude that the loss of KRb molecules with unit
probability is mainly due to the unobserved loss (chemical reactions) compared
to indirect loss and that $y(I) \simeq z(I) \simeq 1$.

An additional possibility occurs for vibrationally excited states of
NaRb~\cite{ye.guo.ea:collisions}.
These molecules experience loss due to complex formation when in their
ground state, but in their first vibrationally excited state they also have
sufficient energy to inelastically de-excite or to chemically react.
In this experiment, neither the inelastic nor the reactive products are observed.
Therefore, inelastic and reactive processes should be regarded as unobserved
absorption processes.
The total loss is therefore described by the absorption coefficient $z(I)$ of
Eq.~\eqref{eq:zformula}.
Here $x(I)$ would characterize the loss due to complex formation, which in the
simplest interpretation can be taken as the same as for the non-reactive ground
state scattering, $x(I)=0.5$.
While $y(I)$ characterizes the losses due to the inelastic and reactive processes.
From the data of the NaRb experiment in $v=1$, the total absorption coefficient
$z(I)=0.93$ has been extracted~\cite{bai.li.ea:model}.
From Eq.~\eqref{eq:zformula} we infer that the unobserved absorption coefficient
is $y(I)=0.8$.
This value, lying close to unity, emphasizes that the unobserved absorption
coefficient, responsible for losses due to inelastic collisions and chemical
reactions, takes a high value close to unity, just as in the KRb case.

\section{Conclusion}
As the  poet decreed, ``Those whom the gods would destroy,
they first make mad'', and so it is for ultracold molecules.
When ultracold molecules collide, a likely outcome is a transformation that
releases energy and sends the molecules fleeing from the trap, effectively
destroying the gas.
But for ultracold alkali dimers this destruction need not be immediate,
if the molecules first, maddeningly, form a collision complex.
Various fates await the molecules upon collisions: elastic scattering,
inelastic scattering, reactive scattering, or resonant complex formation.
In the work we have detailed a simple quantum-defect model capable of treating
all these myriad processes on an equal footing.

In the absence of full scattering matrices, the model captures the essence of
these various processes, providing parameterizations of the various
cross sections.
The model is flexible enough to account explicitly for those processes that
are ultimately observed, and to account implicitly for those that are not.
The result is a framework capable of being adapted to fit the available data
for a given experiment, relating the observables to a small set of
parameters.
These parameters, in turn, represent a tangible goal for microscopic theories
of the four-body dynamics.

The theory as presented treats only the first step of the scattering process,
molecules colliding and heading off on one of the paths, elastic,  inelastic, reactive,
or complex formation.
In particular, the theory does not treat the possible decay of the complex,
to do so will require a more detailed treatment of the decay rate and product
distribution of the complex, work that is currently in progress.

\section*{Acknowledgments}
J. F. E. C. acknowledges that this work was supported by the Marsden Fund of New Zealand (Contract No. UOO1923) and gratefully acknowledges support from the Dodd-Walls Centre for
Photonic and Quantum Technologies.
J. L. B. acknowledges that this material is based upon work supported by the National Science Foundation under Grant Number 1734006, and under grant number 1806971.
G. Q. acknowledges funding from the FEW2MANY-SHIELD Project No. ANR-17-CE30-0015 from Agence Nationale de la Recherche.

\bibliographystyle{naturemag}
\bibliography{../../all}

\end{document}